\documentstyle[12pt]{article}

\def\agt{\buildrel {\mbox{$>$}} \over {\raisebox{-0.8ex}{\hspace{-0.05in}
$\sim$}}}

\def\overlay#1#2{\ifmmode%
\setbox0=\hbox{$#1$}%
\setbox1=\hbox to\wd0{\hss$#2$\hss}\else%
\setbox0=\hbox{#1}%
\setbox1=\hbox to\wd0{\hss#2\hss}\fi%
 #1\hskip-\wd0\box1 }

\begin{document}

\hfill\vbox{\hbox{\bf NUHEP-TH-94-13}\hbox{hep-ph/9406228}
\hbox{June 1994}}\\

\begin{center}
{\Large \bf Studies of Strong Electroweak Symmetry Breaking at Photon
Colliders}\\

\vspace{0.3in}

Kingman Cheung\footnote{Internet address:
{\tt cheung@nuhep.phys.nwu.edu}} \\

\vspace{0.3in}

{\it Dept. of Physics \& Astronomy, Northwestern University, Evanston,
Illinois 60208, USA}
\end{center}

\vspace{0.2in}

\begin{abstract}
It has recently been shown that the studies of strongly-interacting
electroweak symmetry breaking (EWSB) at photon colliders, via photon
splitting into $W$ pair followed by longitudinal $W$-boson scattering, could be
possible.  Here we present a signal-background analysis for the
scattering channels $W^+_L W^-_L \to Z_L Z_L$ and $W^+_L W^-_L \to W^+_L
W^-_L$ with background coming from the standard model (SM) production of
$\gamma\gamma \to WW ZZ$ and $WWWW$, respectively.  We illustrate the analysis
using the SM with a heavy Higgs boson $(m_H\approx 1$ TeV) to
represent a typical strongly-interacting EWSB model and the SM with a light
Higgs boson ($m_H\approx 0.1$ TeV) to represent the background.
We come up with a set of kinematic acceptance to enhance the
signal-to-background ratio.   Extension of the kinematic acceptance to other
strongly-interacting EWSB models is then trivial, and the signal cross
sections  for various EWSB models are calculated.
We found that it is very feasible to probe the EWSB sector at a photon
collider of center-of-mass energy of 2 TeV with a luminosity of just
10~fb$^{-1}$.
\end{abstract}

\thispagestyle{empty}

\newpage
\begin{center}\section{Introduction}
\label{intro}
\end{center}

So far very little is known about the electroweak symmetry-breaking (EWSB)
sector, except that it gives  masses to the vector bosons via spontaneous
symmetry breaking, and also gives masses to fermions via Yukawa couplings.
In the minimal standard model (SM) a  scalar Higgs boson is responsible for
electroweak symmetry-breaking but its mass is not determined by the model.
If in the future no Higgs boson is found below 800 GeV,
the heavy Higgs scenario ($\approx 1$ TeV) will imply a strongly-interacting
Higgs sector because  the Higgs self-coupling $\lambda\sim m_H^2$
becomes strong \cite{quigg}.  However, there is no evidence to favor
models with a scalar Higgs boson, so any models that can break the electroweak
symmetry the same way as the single Higgs boson does can be a candidate for the
EWSB sector.

One of the best ways to uncover the underlying dynamics of the EWSB sector
is to
study the longitudinal vector boson scattering \cite{quigg,chano}.
The Equivalence Theorem \cite{quigg}  recalls, at high energy,
 the equivalence between the longitudinal component $(W_L)$ of the vector
bosons and the corresponding Goldstone bosons ($w$) that were ``eaten"
in the Higgs mechanism.  These Goldstone bosons
originate from the EWSB sector so that their scattering must be via the
interactions of the EWSB sector, and therefore the $W_L W_L$ scattering
can reveal the dynamics  of the EWSB.

Experimentally, the search for the Higgs boson at high energy colliders
are so far all negative.   Probing the EWSB sector at TeV regime
is one of the major goals of all the future
supercolliders.   Ever since the cancellation of the Superconducting
Super Collider, every other opportunity to study EWSB should be
explored.    Recently, the upgraded Tevatron draws some interests in
probing for the Higgs boson.  But studies showed that the machine is
 marginal for discovering the intermediate mass Higgs boson \cite{stange},
 not to mention the heavy Higgs boson or the strong EWSB sector.
The best opportunity will be at the Large Hadron Collider
with a center-of-mass energy 10--14 TeV and a peak
luminosity of $10^{34}\;{\rm cm}^{-2} {\rm s}^{-1}$.  The $e^+e^-$ and $e^-e^-$
machines at 1.5--2 TeV also provide possibilities to probe the strong EWSB
sector.

With the idea of laser backscattering \cite{telnov}
it is relatively inexpensive to convert
a linear $e^+e^-$ or $e^-e^-$ collider into a $\gamma\gamma$ collider.
The resulting photon beams are very monochromatic carrying about 0.8 of the
energy of the parent electron beams.  Also, polarized laser and electron beams
can be employed to further increase the monochromaticity of the photon beam.
For a general review of physics potentials at high energy photon colliders
please refer to Refs.~\cite{brodsky,boudjema}.

Since photon couples to $W$ boson with a coupling strength $g$, same order
as the fermion-$W$ coupling,  we expect the effective $W$ luminosity
inside photons to be of the same order as the luminosity inside electrons
or quarks.  It was shown in Ref.~\cite{hagiwara} that the effective $W$
luminosity inside a photon has a $\log({s/m_W^2})$ enhancement at very high
energy and is  given by
\begin{equation}
\label{lum}
f_{W/\gamma}(x) = \frac{\alpha}{\pi}\left[ \frac{1-x}{x} +\frac{x(1-x)}{2}
\left( \log\frac{s(1-x)^2}{m_W^2} -2 \right) \right ]\,.
\end{equation}
Note that the first term in Eqn.~(\ref{lum}) is very close to the $W$
luminosity inside an electron.

Previously, all studies of EWSB in $\gamma\gamma$ collisions concentrate
on the loop processes $\gamma\gamma \to W_L W_L$ and $Z_L Z_L$ \cite{old}.
Unfortunately, the background from $\gamma\gamma\to W_T W_T$ is almost three
 orders of magnitude larger than the $W_LW_L$ signal.   Although the
signal-to-background ratio  can be improved by requiring the final state
$W$ bosons  away from the beam,  it hardly reduces the $W_TW_T$ background
to the level of the $W_LW_L$ signal.
On the other hand, both the $\gamma\gamma\to Z Z$ signal and background
are  absent on tree level.  But the box diagram contribution to $Z_TZ_T$
has been shown to be very significant at the large $m(ZZ)$ region, so the
$Z_T Z_T$ background is  dominant over the $Z_L Z_L$ signal in the search
of the SM Higgs boson for $m_H\agt 350$ GeV and also in probing  other strong
EWSB signals \cite{ZTZT}.  Unless the polarizations of the final state $ZZ$
and $WW$ pair can be differentiated, it is very hard to use these loop
processes to probe the strong EWSB sector.

It was suggested in Refs.~\cite{brodsky,cheung} that longitudinal
$W$-boson scattering in $\gamma\gamma$ collisions, which is analogous to those
considered at hadronic and $e^+e^-$ colliders, might be useful for probing
the EWSB sector.  The schematic diagram for the longitudinal $W$-boson
scattering is depicted in Fig.~\ref{scheme}.   An advantage of this process
is that tagging the spectator $W$ bosons, in addition to the strongly-scattered
vector bosons, can eliminate all the backgrounds from
$\gamma\gamma \to W_T W_T$ and $Z_T Z_T$.
It was also shown in Ref.~\cite{cheung}
that the cross sections for the signal of various strongly-interacting
EWSB models are large enough to be observable.
But at that time, backgrounds from the SM have not been calculated, so it is
not possible to draw any conclusions.  Nevertheless,
according to a preliminary result \cite{jikia}, the SM backgrounds from
$\gamma\gamma \to WWZZ$ and $WWWW$ are
manageable with respect to the signals.  It is the purpose of this paper to
investigate  independently the possibility.

The calculation of signals for various models has been given in
Ref.~\cite{cheung}, in which the method of effective $W$ luminosity inside
a photon is used.     This method has a disadvantage that
the kinematics of the spectator $W$ bosons cannot be calculated exactly,
so any acceptance cuts on the spectator $W$ bosons are unrealistic.
The way we do here is to carry out exact calculations for the processes
$\gamma\gamma \to WWZZ$ and $WWWW$.
The heavy Higgs boson, which is considered as a typical strong EWSB model,
can be incorporated consistently into the SM by putting the Higgs-boson mass
$m_H$ very large, say 1 TeV, and therefore  it can be calculated exactly;
while for other EWSB models
we still have to rely on the method of effective $W$ luminosity, in which
the luminosity function is folded with the subprocess cross section.
But based on the fact that the kinematics of the spectator $W$ bosons
is insensitive to different strong EWSB models, we expect that
the tagging efficiencies of the spectator $W$ bosons,
which we can obtain consistently for the heavy Higgs model from  the exact
calculations of $\gamma\gamma \to WWZZ$ and $WWWW$, can be applied trivially to
other strong EWSB models.  In Secs.~\ref{II} and \ref{III}, we
confront the heavy Higgs-boson signal against the SM background in the
channels $\gamma\gamma \to WWZZ$ and $WWWW$, respectively.  We come
up with a favorable set of acceptance cuts to enhance the signal-to-background
ratio and also obtain the tagging efficiencies for the spectator $W$
bosons.  We then apply these tagging efficiencies to other EWSB
models in Sec.~\ref{IV}.

Before we proceed, let us define the signal and background more precisely.
The background is essentially the expectation from the SM with a very light
Higgs boson.  Cross sections of the processes $\gamma\gamma \to WWZZ$ and
$WWWW$ with a light Higgs boson ($m_H=0.1$ TeV) then represent the
backgrounds.
On the other hand, the signal can be considered as enhancement over the SM
expectation.  As we mentioned before, the calculation of the heavy
Higgs-boson signal can be incorporated consistently into the SM
by putting $m_H$ very large, say 1 TeV.    The signal is then
defined as the difference between the following cross sections
$\sigma(m_H=1$ TeV) $-\;\sigma(m_H=0.1$ TeV).
Signal for other models is calculated by folding the
subprocess cross sections $\hat \sigma (W^+_L W^-_L\to Z_L Z_L,\,W^+_L W^-_L)$
with the effective $W$ luminosity inside a photon, which is given
in Eqn.~(\ref{lum}).

We first concentrate on  the channel $\gamma\gamma \to WWZZ$ because it is
relatively simple in the sense that the $ZZ$ pair must come from the
longitudinal $W$-boson  scattering while the final state $W$ bosons are  the
spectators.  It is therefore straightforward
to implement the acceptance cuts on the  strongly-scattered $Z$ bosons and
on the spectator $W$ bosons, separately.  However, for
the channel $\gamma\gamma \to WWWW$ it is more complicated to implement
the kinematic cuts because it is ambiguous to determine which $W$ bosons come
out from the strong scattering region and which $W$ bosons are the
spectators.   We adopt the following procedures.   We reorder the $W$
bosons according to the absolute values of their rapidities.
Those two with smallest absolute rapidities are the bosons
coming  out from the strong scattering region, while
those two with largest absolute rapidities are the spectators.

The organization is as follows.  In the next section, we present the
signal-background analysis for the channel $\gamma\gamma\to WWZZ$.
In Sec.~\ref{III}, we repeat the same analysis for the channel $\gamma\gamma
\to WWWW$. In Sec.~\ref{IV}, we calculate the signal for various
strongly-interacting  EWSB models.
We  reserve Sec.~\ref{V} for discussions and conclusions.

\begin{center}
\section{$\gamma\gamma \to WW ZZ$}
\label{II}
\end{center}

We illustrate in this section the signal-background analysis for the
channel $\gamma\gamma \to WWZZ$,  with the signal of a 1 TeV Higgs boson
defined by
\begin{equation}
\sigma(m_H=1\;{\rm TeV}) \quad - \quad \sigma(m_H=0.1\;{\rm TeV})
\end{equation}
and the background is represented by $\sigma(m_H=0.1$ TeV).
Typical Feynman diagrams for the process $\gamma\gamma\to WWZZ$ are shown
in Fig.~\ref{feynman}.  The complete set of Feynman diagrams contains the heavy
Higgs-boson signal that we are considering ({\it e.g.}, in
Fig.~\ref{feynman}(a)).   This is the reason why we said above that the
heavy Higgs-boson signal can be incorporated consistently into the SM.
We use the package MADGRAPH \cite{tim} to generate the complete set of Feynman
diagrams and the fortran code for the squared amplitude.  Totally, there are
74 Feynman diagrams in the unitary gauge.
We present the total cross sections for the process $\gamma\gamma\to WWZZ$
versus the center-of mass energies of the $\gamma\gamma$ system with
$m_H=0.1$ and 1 TeV in Fig.~\ref{cross}.  Enhancement of the total cross
section due to the heavy-Higgs-boson  exchange is only significant for
$\sqrt{s_{\gamma\gamma}} \agt 1.5$ TeV.  Therefore, in the following we choose
$\sqrt{s_{\gamma\gamma}}=2$ TeV to illustrate the confrontation of the heavy
Higgs-boson signal against the background.  Later, we also show the
results for other center-of-mass energies.

We will look at some kinematic variables to enhance the
signal-to-background ratio.  Thanks to some intensive studies of longitudinal
vector boson scattering at hadronic supercolliders \cite{bagger,hadron},
we can borrow their strategies.   The strongly-scattered $Z_L$ bosons
should have larger transverse momentum and larger invariant mass $m(ZZ)$ in
the central rapidity region than the $Z$ bosons  from the
background; while the spectator $W$ bosons, coming from the photon splitting,
tend to be in the forward rapidity region.
Therefore, we begin with the acceptance cuts
\begin{equation}
\label{zz}
m(ZZ) > 500\; {\rm GeV} \quad {\rm and} \quad    |y(Z)| < 1.5
\end{equation}
on the strongly-scattered $ZZ$ pair, and the basic acceptance to tag both
spectator $W$ bosons:
\begin{equation}
\label{W}
p_T(W)>25 \; {\rm GeV}  \quad {\rm and} \quad   |y(W)| < 3 \;.
\end{equation}
These  spectator $W$ bosons have to be tagged in order to eliminate the
$\gamma\gamma \to W_T W_T$ or $Z_T Z_T$
backgrounds.    We use a wide rapidity coverage of 3 because we expect that
the spectator $W$ bosons for the signal are very forward  but they
can hardly go beyond $|y(W)|=3$ in rapidity at
$\sqrt{s_{\gamma\gamma}}=2$ TeV, as indicated in Fig.~\ref{rapid}.
To demonstrate the fact that the spectator
$W$ bosons for the signal are more forward than those for the background,
we show the rapidity distribution of the more-forward $W$ boson for the case
of $m_H=1$ TeV and for the background ($m_H=0.1$ TeV) in Fig.~\ref{rapid}.
{}From the figure it is advantageous to require {\it at least one} of the
spectator $W$ bosons  in the forward rapidity region defined by
\begin{equation}
\label{rap}
1.5 < |y(W)| < 3.0\;.
\end{equation}
We also look at the transverse momentum $p_T$ distribution of the $Z$ bosons,
as we expect that the strongly-scattered $Z$ bosons should have larger
$p_T$ than the $Z$ bosons from the background.
We show the distribution of the min$(p_T(Z_1),p_T(Z_2))$
in Fig.~\ref{pt}. From the figure, a $p_T$ cut of
\begin{equation}
\label{ptcut}
p_T(Z) > 250 \;{\rm GeV}
\end{equation}
can further improve the signal-to-background ratio.
We summarize in Table~\ref{table1} the cross sections for  various
combinations of the cuts in Eqns.~(\ref{zz}), (\ref{W}), (\ref{rap}), and
(\ref{ptcut}).
In fact, we can gain in the signal-to-background ratio by tightening
the $p_T(Z)$ cut or by imposing other cuts, {\it e.g.} $\Delta p_T(ZZ)=
|\vec p_T(Z_1) - \vec p_T(Z_2)|>600$ GeV, but at the same time
we are losing signal events.  We also show in Table~\ref{table1} the
significance of the signal defined by $S/\sqrt{B}$, where $S$ and $B$ are the
number of signal and background events with an integrated  luminosity of
10~fb$^{-1}$.

Next, we estimate the tagging efficiencies for the spectator $W$ bosons.  The
acceptance cuts on the spectators are given in Eqns.~(\ref{W}) and
(\ref{rap}).   The efficiency can be calculated from Table~\ref{table1}.
The last second row shows the cross sections with all acceptance cuts
imposed; while the last row shows the cross sections with the acceptance cuts
on $Z$ bosons only.   The tagging efficiency of the spectator $W$ bosons
for the signal is then
\begin{equation}
\frac{\sigma({\rm signal})|_{\mbox{\scriptsize last second row}}}
     {\sigma({\rm signal})|_{\mbox{\scriptsize  last row}}}
\;=\;\frac{10.8\;{\rm fb}}{13.6\;{\rm fb}} \;=\;  79 \%\;.
\end{equation}
This efficiency is applied in Sec.~\ref{IV} to estimate the cross sections for
other strong EWSB models.

\begin{center}
\section{$\gamma\gamma \to WWWW$}
\label{III}
\end{center}

The complete set of Feynman diagrams and the fortran code for the squared
amplitude are also generated by MADGRAPH \cite{tim}.  There are totally
240 contributing Feynman diagrams in the unitary gauge.
As explained in the Introduction, this channel is more complicated because of
various combinations.  It can have  enhancement from the strong scattering
channels $W^\pm_L W^\pm_L \to W^\pm_L W^\pm_L$ and $W^+_L W^-_L \to W^+_L
W^-_L$.  According to Ref.~\cite{cheung}, the signal for the like-charge
channels is substantially smaller than the signal for the opposite-charge
channel.   The dominance of the opposite-charge channel over the like-charge
channels is due to the presence of a s-channel resonance in the
opposite-charge channel and the absence of any doubly-charged  resonance
in the like-charge channels.    Therefore,  for the following we only
concentrate on the opposite-charge scattering channel.

Since we are going to impose very different acceptance cuts on the
strongly-scattered $W$ bosons and the spectator $W$ bosons, we have to
distinguish them.  As mentioned in the Introduction, we reorder the absolute
rapidities of the $W$ bosons according to
$|y(W_1)|<|y(W_2)|<|y(W_3)|<|y(W_4)|$.  We then assume the first
two $W$ bosons with
smallest absolute rapidities to be the strongly-scattered $W$ bosons as we
expect them to be central; while the last two $W$ bosons with largest absolute
rapidities to be the spectators as we expect them to be forward.
We proceed closely as in Sec.~\ref{III}.  We impose the acceptance cuts
\begin{equation}
\label{ww}
m(WW) > 500\; {\rm GeV} \quad {\rm and} \quad |y(W)|<1.5
\end{equation}
on the strongly-scattered $W$ bosons, and the basic cuts
\begin{equation}
\label{W-w}
p_T(W_{\rm sp}) > 25\; {\rm GeV} \quad {\rm and} \quad |y(W_{\rm sp})|<3
\end{equation}
to tag both spectator $W$ bosons.  We put a ``sp" in the subscript to
indicate that they are the spectator $W$ bosons.    We also require at
least one forward spectator $W$ boson in the rapidity region defined by:
\begin{equation}
\label{rap-w}
1.5 < |y(W_{\rm sp}) | < 3\,,
\end{equation}
since we expect that the behavior of the spectator $W$ bosons here is the same
as the spectator $W$ bosons in the $ZZ$ channel.
Then we looked at the $p_T$ distribution of the strongly-scattered $W$
bosons to determine the value needed to further suppress the background and
we have chosen
\begin{equation}
\label{ptcut-w}
p_T(W) > 250 \;{\rm GeV}\,.
\end{equation}
We summarize the cross sections for  various combinations of the cuts in
Table~\ref{table2}, which is similar to Table~\ref{table1}.
The tagging efficiency of the spectator $W$ bosons for this channel is
\begin{equation}
\frac{\sigma({\rm signal})|_{\mbox{\scriptsize  last second row}} }
     {\sigma({\rm signal})|_{\mbox{\scriptsize  last row}} }
\,=\, \frac{20.2\;{\rm fb}}{25.6\;{\rm fb}} \,=\, 79\%\;,
\end{equation}
which happens to be  the same as the $ZZ$ channel within the first two
significant digits.

\begin{center}
\section{Signal for Strong EWSB Models}
\label{IV}
\end{center}

Description of some strongly-interacting EWSB models and the amplitude
function predicted by each of the models can be found in Ref.~\cite{bagger}.
Here we calculate the signal cross sections for various models by the method
of effective $W$ luminosity, in addition to the heavy
Higgs-boson model that we have studied in the Secs.~\ref{II} and \ref{III}.
Each of the $W_L W_L$ scattering amplitudes grows with energy until reaching
the resonances, {\it e.g.} a technirho.    The presence of
the resonances (scalar or vector) is the natural unitarization to the
scattering amplitudes, except that there might be slight violation of
unitarity around the resonance peak.  After the resonance, the scattering
amplitudes  will stay below the unitarity limit.    The models can be
classified according to the spin and isospin properties of the resonance
fields,  which are to unitarize the $W_L W_L$ scattering
amplitudes.   There are scalar-like, vector-like, and nonresonant models.
For scalar-like models we employ the SM with a 1 TeV Higgs boson,
the model with a chirally-coupled scalar of mass  $m_S=1$~TeV and
width $\Gamma_S=350$~GeV, and $O(2N)$ model with the cutoff $\Lambda=2$~TeV.
For the vector-like models we choose the chirally-coupled vector field
(technirho) of masses $m_\rho=1.2$ and 1.5~TeV, and $\Gamma_\rho=0.5$
and 0.6~TeV, respectively.
In the extreme case of no light resonance (nonresonant model),  unitarity
is likely to be saturated before reaching the lightest resonance.  Here we
employ the Low Energy Theorem (LET)-derived  amplitude function,
$A(s,t,u)=s/v^2$, for the nonresonant model and extrapolate it to high
energy.  We might have to worry about unitarity violation in the scattering
amplitudes.  Let us take a look at the LET-derived amplitude.   From the
partial wave analysis, the only nonzero partial wave coefficients
$a^I_J$ are $a^0_0$, $a^1_1$, and $a^2_0$.  Among the nonzero $a^I_J$'s,
$a^0_0$ saturates the unitarity ($|a^I_J|<1$) at the lowest energy
$4\sqrt{\pi}v\approx 1.7$~TeV, which is the center-of-mass energy of the $W_L
W_L$ system.     So for $\gamma\gamma$ colliders of 1.5--2 TeV
that we are considering, unitarity violation
should not happen, and therefore we simply extrapolate the LET amplitudes
without any unitarization.  Later, we also extend the results to
$\sqrt{s_{\gamma\gamma}}=3$ TeV.
But for simplicity we leave out the unitarization procedures
so that our results for $\sqrt{s_{\gamma\gamma}}\agt 2$ TeV might
slightly over-estimate the actual cross sections, or in other words, they
represent some upper bounds for the cross sections.

Before we present the results for the signals, let us examine the validity of
the method of effective $W$ luminosity by comparing the heavy Higgs-boson
signal obtained by the exact calculation and by the method of effective
$W$ luminosity.   In Secs.~\ref{II} and \ref{III},
we already have the results for the exact calculations of the 1 TeV
Higgs-boson signal at $\sqrt{s_{\gamma\gamma}}=2$ TeV, which are
listed in Tables~\ref{table1} and \ref{table2}.  We first compare the
$Z_L Z_L$ channel.  Using the method of effective $W$ luminosity the signal
is given by
\begin{equation}
\sigma(s_{\gamma\gamma}) = \int dx_1 dx_2\; f_{W/\gamma}(x_1)
f_{W/\gamma}(x_2)\; \hat \sigma(W^+_L W^-_L \to Z_L Z_L, \hat s =x_1 x_2
s_{\gamma\gamma})\;,
\end{equation}
where $f_{W/\gamma}(x)$ is the effective $W$ luminosity inside a photon
given in Eqn.~(\ref{lum}) and $\hat \sigma(\hat s)$ is given  by
\begin{equation}
\hat \sigma (\hat s) = \int d(PS)\; \frac{1}{2 \lambda^{1/2}(\hat s, m_W^2,
m_W^2)} \left| {\cal M}(W^+_L W^-_L \to Z_L Z_L) \right |^2 \;,
\end{equation}
where $\lambda(x,y,z)=(x^2+y^2+z^2-2xy-2yz-2zx)$ and $PS$ is the phase space
factor including  the symmetry factor of 1/2.   The scattering
amplitude ${\cal M}$ is written as
\begin{equation}
{\cal M}(W^+_L W^-_L \to Z_L Z_L) = A(\hat s,\hat t,\hat u)\;,
\end{equation}
where $\hat s,\hat t,$ and $\hat u$ in the above equation refer to the
$W^+_L W^-_L$ system and $A(\hat s,\hat t,\hat u)$ is the amplitude
function predicted by each strong EWSB model.
For the heavy Higgs-boson model $A(\hat s,\hat t,\hat u)$ is given by
\begin{equation}
A(\hat s,\hat t,\hat u) = \frac{-m_H^2}{v^2} \, \left( 1+
\frac{m_H^2}{\hat s-m_H^2 + i m_H \Gamma_H \theta(\hat s)} \right )\;,
\end{equation}
where $v \approx 246$ GeV and $\theta(\hat s)=1\,(0)$ for $\hat s>0$
(otherwise).  $\Gamma_H$ is the decay width of the Higgs boson and we take
$\Gamma_H=0.5$ TeV for $m_H=1$ TeV.
With only the acceptance cuts in Eqns.~(\ref{zz}) and (\ref{ptcut})
on the strongly-scattered $ZZ$ pair  at
$\sqrt{s_{\gamma\gamma}}=2$ TeV, the method of
effective $W$ luminosity gives $\sigma(\gamma\gamma \to W_{\rm sp}W_{\rm sp}
 Z_L Z_L)=12.7$ fb,
which is within  7\%  of  the result (13.6 fb) for the exact calculation.
For the $W^+_L W^-_L$ channel we do similar comparison.  The scattering
amplitude ${\cal M}(W^+_L W^-_L \to W^+_L W^-_L)$ is again expressed in term
of the amplitude function:
\begin{equation}
{\cal M}(W^+_L W^-_L \to W^+_L W^-_L) = A(\hat s,\hat t,\hat u) +
A(\hat t,\hat s,\hat u) \;.
\end{equation}
With the acceptance cuts in Eqns.~(\ref{ww}) and (\ref{ptcut-w})
on the strongly-scattered $WW$ pair at
$\sqrt{s_{\gamma\gamma}}=2$ TeV, the method of effective $W$ luminosity gives
$\sigma(W^+_L W^-_L \to W^+_{\rm sp}  W^-_{\rm sp} W^+_L W^-_L) = 29.0$ fb,
which is still within 15\% of the result (25.6 fb) for  the exact calculation.
Therefore, we have justified here the validity of the method of
 effective $W$ luminosity with the acceptance cuts: $m(WW/ZZ)>500$ GeV,
$|y(W/Z)|<1.5$, and
$p_T(W/Z)>250$ GeV on the strongly-scattered $W/Z$ bosons.  We do expect that
the approximation works better at very large invariant mass and central
rapidity phase space region.  From now on, we  use the method of effective
$W$ luminosity to calculate the signal for various models including
the heavy Higgs boson.

With the 79\% tagging efficiency of the spectator $W$ bosons
for both $W^+_L W^-_L \to Z_L Z_L$ and
$W^+_L W^-_L \to W^+_L W^-_L$ channels, we show the cross sections of the
signal for various models at $\sqrt{s_{\gamma\gamma}}=2$ TeV and the
significance of each in Table~\ref{table3}.   From Table~\ref{table3},
we can see that both channels are very sensitive to the presence of
scalar-like resonances, but $W^+_L W^-_L$ channel is far more sensitive to the
presence of vector-like resonances than the $Z_L Z_L$ channel.  On the other
hand, in the extreme case of no light resonances, the $Z_L Z_L$ channel
is enhanced more than the $W^+_L W^-_L$ channel by about 50\%.

Although the energy of photon colliders is limited by the Next Linear Collider
designs, it is still instructive to show the signal and background cross
sections  at other
center-of-mass energies.  However, there is a technical difficulty that the
tagging efficiency for the spectator $W$ bosons by the cuts in
Eqns.~(\ref{W}) and (\ref{rap}) is likely to
vary with the center-of-mass energies.    From Fig.~\ref{rapid}, we know that
there are hardly any signal and background events beyond $|y(W_{\rm sp})|=3$
at $\sqrt{s_{\gamma\gamma}}=2$ TeV, due to the finite $W$-boson mass.
However, we do expect that for $\sqrt{s_{\gamma\gamma}}>2$ TeV the spectator
$W$ bosons can go further out in the forward rapidity region and we verified
that at $\sqrt{s_{\gamma\gamma}}=3$ TeV the signal events can go up to about
$|y(W_{\rm sp})|=3.5$ with a substantial number of them beyond
$|y(W_{\rm sp})|=3$; while the majority of the background events are still
within $|y(W_{\rm sp})|<3$.   Therefore, in order to
maintain a large ($\approx 80$\%) tagging  efficiency at
$\sqrt{s_{\gamma\gamma}}=3$ TeV, we have to
extend the forward rapidity coverage from 3 to 3.5, while such an extension in
the rapidity coverage should not affect significantly the background cross
section since the majority of the background events are within $|y(W_{\rm
sp})|<3$.   Therefore, we expect that at different $\sqrt{s_{\gamma\gamma}}$
we have to adjust the rapidity coverage for the spectator $W$ bosons in order
to maintain a large tagging efficiency.
Instead of presenting our summary curves with different cuts at
different center-of-mass energies, we adopt the following, for simplicity.
We calculate the background for $\sqrt{s_{\gamma\gamma}}=1-3$ TeV with the
same acceptance cuts: $m(ZZ/WW)>500$ GeV, $|y(Z/W)|<1.5$, and $p_T(Z/W) >250$
GeV on the strongly-scattered $Z/W$ bosons, and $p_T(W_{\rm sp})>25$ GeV,
$|y(W_{\rm sp})|<3$, and requiring  at least one forward spectator $W$ boson
in the range $1.5<|y(W_{\rm sp})|<3$ for the spectator $W$ bosons.
Whether or not  extending the rapidity coverage  should not change the
background cross sections significantly since the majority of the background
events are within $|y(W_{\rm sp})|<3$.
On the other hand, the cross sections  for various models at
$\sqrt{s_{\gamma\gamma}}=1-3$ TeV are calculated with only the acceptance
cuts on the strongly-scattered $W/Z$ bosons and then multiplied by a
constant 79\% tagging efficiency  to represent the effect of tagging the
spectator $W$ bosons.  Although the 79\% tagging efficiency is only valid for
$\sqrt{s_{\gamma\gamma}}=2$ TeV with the rapidity coverage up to 3, we do
expect that similar tagging efficiencies can be obtained by extending the
rapidity coverage, which depends on $\sqrt{s_{\gamma\gamma}}$, from 3
gradually to 3.5 at $\sqrt{s_{\gamma\gamma}}=3$ TeV; while this extension in
rapidity coverage should not affect the background cross sections
significantly.
The cross sections for various models in the channels $W^+_L W^-_L \to Z_L
Z_L$ and $W^+_L W^-_L \to W^+_L W^-_L$ are shown in Fig.~\ref{summary1}(a) and
(b), respectively.  The backgrounds from the SM production of $\gamma\gamma
\to WWZZ$ and $WWWW$ with $m_H=0.1$ TeV are also shown.  The 79\%  tagging
efficiency of the spectator $W$ bosons for both $Z_L Z_L$ and $W^+_L
W^-_L$ channels has been multiplied in the signal curves of
Figs.~\ref{summary1}(a) and (b).    From Figs.~\ref{summary1}(a) and (b),
the $Z_L Z_L$ channel seems doing better
than the $W^+_L W^-_L$ channel, as the $\gamma\gamma\to WWZZ$
background can be suppressed
below most of the signal curves except for the models of heavy vector
resonance and of LET.    This is due to the fact that the cross section for
$\gamma\gamma\to WWWW$ receives many contributions that are not sensitive to
the Higgs-boson mass.  Obviously, the higher the center-of-mass energies, the
better is the possibility of probing the strongly-interacting
 EWSB scenario.  At
$\sqrt{s_{\gamma\gamma}}=1.5$ TeV, although the signal-to-background ratio is
greater than 1 for both channels and for most of the models, the number of the
signal events  might be too small for any practical observation, unless a very
high luminosity  can be achieved, say 100~fb$^{-1}$ \cite{cheung}.
For $\sqrt{s_{\gamma\gamma}}\agt 2$ TeV the cross sections for the signal
are much larger and a large signal-to-background ratio is still maintained,
so the feasibility to probe the EWSB improves significantly.
According to Table~\ref{table3}, a center-of-mass energy
$\sqrt{s_{\gamma\gamma}}$ of 2 TeV with an integrated luminosity of
10~fb$^{-1}$ is already sufficient to probe the strong EWSB scenario.
However, for the present highest energy $e^+e^-$ collider
designs of 1.5 TeV it can at most be converted to a photon collider of
energy about 1.2 TeV and this is certainly not enough to probe the EWSB at
TeV regime.

\begin{center}
\section{Discussions}
\label{V}
\end{center}

So far we have peformed the signal-background analysis with the assumptions of
a perfect monochromatic $\gamma\gamma$ collider, and ignoring any QCD-related
backgrounds and the decays of the vector bosons.  We are going to
discuss them in order.    First, we discuss the decays of the vector bosons.
Since it is necessary to identify the $W$ and $Z$ bosons, only the decay
modes, in which the $W$ and $Z$ bosons can be fully reconstructed, are
considered.   Therefore, for the $W$ bosons coming out from the
strong-scattering region it has to decay hadronically; while the
$Z$ bosons can decay into hadrons and leptons.
The combined branching ratio for the
strongly-scattered $WW$ pair is [Br($W\to q\bar q')]^2 =(0.7)^2 \approx 0.5$;
while that for the $ZZ$ pair is [Br($Z\to q\bar q,\,\ell\bar \ell)]^2 =(0.8)^2
\approx 0.6$.    Furthermore, we also have to identify the spectator $W$ bosons
with full reconstruction so as to eliminate the QCD backgrounds.  The branching
ratio for the spectator $W$ bosons is then $(0.7)^2 \approx 0.5$.   In total,
we have a combined branching ratio of 25\% (30\%) for the $W^+_L W^-_L$
($Z_L Z_L$)  channel.   It implies that after we take into account of the
decay branching ratios  the signal and background cross
sections in Table~\ref{table3} are quartered and the significance of the
signal is halved, which does not affect our conclusion that a 2 TeV photon
collider with a luminosity of 10~fb$^{-1}$ is sufficiently feasible to
probe the EWSB sector.
QCD backgrounds should not be serious since we always require to
identify the $W$ and $Z$
bosons by fully reconstructing their masses.  By reconstruction, most of the
backgrounds from QCD production of jets are eliminated.  Production of
$\gamma\gamma\to t\bar t t\bar t$ might be a possible background, but the
presence of the $b$-jets and the top-mass reconstruction can help eliminating
this background.

A perfect monochromatic $\gamma\gamma$ collider might be possible in the
future but even the best up-to-date design, the laser
backscattering \cite{telnov},   cannot produce  perfect monochromaticity.
The energy spectrum of the  photon beam with respect to the
parent electron beam is continuous with a peak at $x\approx 0.83$.  Therefore,
being more realistic we present the summary curves again but folded with the
the luminosity of the photon beam  obtained from  laser backscattering.  The
luminosity function for the photon spectrum using unpolarized laser and
electron beams is given by \cite{telnov}
\begin{equation}
\label{lumgg}
f_{\gamma /e}(x) = \frac{1}{D(\xi)} \left[ 1-x +\frac{1}{1-x}
-\frac{4x}{\xi(1-x)} + \frac{4x^2}{\xi^2 (1-x)^2} \right] \,,
\end{equation}
where
\begin{equation}
\label{D_xi}
D(\xi) = (1-\frac{4}{\xi} -\frac{8}{\xi^2}) \ln(1+\xi) + \frac{1}{2} +
\frac{8}{\xi} - \frac{1}{2(1+\xi)^2}\,,
\end{equation}
with $\xi =\frac{4E_0 \omega_0}{m_e^2}$, $x=\omega/E_0$, $\omega_0$ is the
energy of the laser photon, $E_0$ is the parent electron beam energy, and
$\omega$ is the energy of the converted photon.    The allow range of $x$ is
$0 \le x \le x_{\rm max}=\xi/(1+\xi)$.  $\xi$ is chosen to be 4.8 in order to
avoid the electron-positron pair creation from the fusion of the laser
photon and the converted photon.  Once $\xi$ is chosen, everything is fixed.
The luminosity function is folded with the subprocess cross sections as
follows:
\begin{equation}
\sigma(s) = \int dx_1 dx_2 \,f_{\gamma/e}(x_1) f_{\gamma/e}(x_2)  \int dx_3
dx_4\, f_{W/\gamma}(x_3)f_{W/\gamma}(x_4)\; \hat \sigma (W^+_L W^-_L \to Z_L
Z_L,\,W^+_L W^-_L )\;
\end{equation}
to obtain the cross sections for the signal at the center-of-mass energy
$\sqrt{s}$ of the parent $e^+e^-$ collider; while the cross section for the
SM background is
\begin{equation}
\sigma(s) = \int dx_1 dx_2 \,f_{\gamma/e}(x_1) f_{\gamma/e}(x_2)  \;
\hat \sigma (\gamma\gamma \to WWZZ,\,WWWW )\;.
\end{equation}
The results are presented in Fig.~\ref{summary2}(a) and (b), respectively, for
the $e^+e^- \stackrel{\rm laser}{\to} \gamma\gamma \to WWZZ$ and $WWWW$ as a
function of the center-of-mass energies of the parent $e^+e^-$ collider.
We have multiplied to the signal curves a constant 79\% tagging
efficiency to represent the effect of tagging the spectator $W$ bosons, and
the background curves are calculated exactly with all the acceptance cuts.
The shape and the relative size of the signal and background curves
do not change significantly from
Figs.~\ref{summary1} to Figs.~\ref{summary2}, in which the photon
spectrum is folded.  However, the actual values of the cross sections drop
substantially, indicating that the monochromaticity of the photon beams is
very important.  This is easy to understand that the cross section of the
signal increases quite sharply with $\sqrt{s_{\gamma\gamma}}$, as demonstrated
in Figs.~\ref{summary1}, and therefore the middle to lower end of the photon
spectrum by laser backscattering can hardly contribute to the strong
EWSB signal.    Since only the upper end of the photon spectrum can contribute
to the signal, it is necessary to use polarized laser and electron beams to
increase the monochromaticity of the colliding photon beams \cite{telnov}.
If the monochromaticity of the photon beam can approach the limit of being
perfect about 0.8 of the parent electron-beam energy, an $e^+e^-$ machine of
2.5 TeV would be sufficient to be converted into a 2 TeV photon collider,
which has been concluded, in the last section, feasible to probe the strong
EWSB scenario.  We will not comment any more on what energy of an $e^+e^-$
collider operating in the $\gamma\gamma$ mode is enough to probe the EWSB
sector, but only emphasize that $\sqrt{s_{\gamma\gamma}}=2$ TeV with a
luminosity of 10~fb$^{-1}$ is sufficient.

We have presented a signal-background analysis for studying the strong $W_L
W_L$ scattering at $\gamma\gamma$ colliders.  We confront the signal of
various strong EWSB models against the worst irreducible background from the SM
production of $\gamma\gamma \to WWZZ$ and $WWWW$.  We have demonstrated, with
the analysis on the vector-boson level, that with our acceptance cuts
the background can be substantially reduced to a level smaller than the signal,
and the signal still maintains a very  high significance with 10 fb$^{-1}$
luminosity.   In principle, a complete Monte Carlo simulation including the
decays of the $W/Z$ bosons, the smearing of the momentum of the decay
products, and the true detector acceptance  is needed to establish
the viability.  Nevertheless, we have shown, as a first step,  that it is
very feasible to probe the EWSB by studying the longitudinal $W$-boson
scattering (Fig.\ref{scheme}) at photon colliders, provided that the
center-of-mass energy of the $\gamma\gamma$ system is of the order 2 TeV with
a luminosity of just 10~fb$^{-1}$, or provided that the center-of-mass energy
 is 1.5 TeV but with a high luminosity of the order 100~fb$^{-1}$.

\bigskip

\section*{Acknowledgements}

Special thanks to Tim Stelzer for providing the $\gamma\gamma \to WWWW$ code
and for useful discussions on the package MADGRAPH.
Also thanks to Bill Long for information on MADGRAPH.
This work was supported by the U.~S. Department of Energy,
Division of High Energy Physics, under Grant DE-FG02-91-ER40684.
%-----------------------------------

%%%%%%%%%%%%%%%
\begin{table}[p]
\caption[]
{\label{table1}Table showing cross sections (fb) for various combinations of
the acceptance cuts in Eqn.~(\ref{zz}), (\ref{W}), (\ref{rap}), and
(\ref{ptcut})
for the channel $\gamma\gamma\to WWZZ$ at $\sqrt{s_{\gamma\gamma}}=2$ TeV.
In the fourth column the signal is defined as $\sigma(m_H=1$ TeV)
$-\,\sigma(m_H=0.1$ TeV).  The last column shows the significance $S/\sqrt{B}$
of the signal with an integrated luminosity of 10 fb$^{-1}$.
 }
\centering
\medskip
\begin{tabular}{|@{\extracolsep{0.25in}}c|c|c|c|c|}
\hline
\hline
& $\sigma(m_H=1$ TeV) & $\sigma(m_H=0.1$ TeV) &  Signal & $S/\sqrt{B}$\\
\hline
No cuts & 92.4 & 64.4 & 28.0 & 11   \\
(\ref{zz})+(\ref{W}) & 33.7 & 17.7 & 16.0 & 12 \\
(\ref{zz})+(\ref{W})+(\ref{rap}) & 24.8 & 10.4 & 14.4 & 14 \\
(\ref{zz})+(\ref{W})+(\ref{rap})+(\ref{ptcut}) & 14.5 & 3.7 & 10.8 & 18 \\
(\ref{zz})+(\ref{ptcut}) & 21.8 & 8.2 & 13.6 & 15 \\
\hline
\end{tabular}
\end{table}

%%%%%%%%%%%%%%%%%%%%%%%%%%%%%%%%%%%%%%%%%%
\begin{table}[p]
\caption[]
{\label{table2}Table showing cross sections (fb) for various combinations of
the acceptance cuts in Eqn.~(\ref{ww}), (\ref{W-w}), (\ref{rap-w}), and
(\ref{ptcut-w}) for the channel $\gamma\gamma\to WWWW$ at
$\sqrt{s_{\gamma\gamma}}=2$ TeV.  In the fourth column the signal is defined
as $\sigma(m_H=1$ TeV) $-\,\sigma(m_H=0.1$ TeV).   The last column shows the
significance $S/\sqrt{B}$  of the signal with an integrated luminosity of
10 fb$^{-1}$.
 }
\centering
\medskip
\begin{tabular}{|@{\extracolsep{0.25in}}c|c|c|c|c|}
\hline
\hline
& $\sigma(m_H=1$ TeV) & $\sigma(m_H=0.1$ TeV) &  Signal & $S/\sqrt{B}$\\
\hline
No cuts & 311  & 264  & 47  & 9.1  \\
(\ref{ww})+(\ref{W-w}) & 80.0 & 51.9  & 28.1  & 12 \\
(\ref{ww})+(\ref{W-w})+(\ref{rap-w}) &  64.4 & 38.9 & 25.5  & 13 \\
(\ref{ww})+(\ref{W-w})+(\ref{rap-w})+(\ref{ptcut-w}) & 34.9  & 14.7 & 20.2 &
17\\
(\ref{ww})+(\ref{ptcut-w}) & 50.4  & 24.8  & 25.6  & 16  \\
\hline
\end{tabular}
\end{table}

%%%%%%%%%%%%%%%%%%%%%%%%%%%%%%%%%%%%%%%%%%%%%%%%%%%%%%%%%%%%%%%%%%%%
\begin{table}[p]
\caption[]
{\label{table3}
Table showing cross sections (fb) for various EWSB models and background at
$\sqrt{s_{\gamma\gamma}}=2$ TeV.   The background is calculated exactly with
the full set of acceptance cuts, and the signal is calculated by the method of
effective $W$ luminosity with only the acceptance cuts on the
strongly-scattered $WW/ZZ$ pair and then multiplied by a constant 79\%
tagging efficiency to represent the effect of tagging the spectator $W$ bosons.
The significance is calculated with a luminosity of 10 fb$^{-1}$.
}
\centering
\bigskip
\begin{tabular}{|l@{\extracolsep{0.25in}}|cc|cc|}
\hline
\hline
&  $\sigma(Z_L Z_L$) & $S/\sqrt{B}$  & $\sigma(W^+_L W^-_L$) & $S/\sqrt{B}$ \\
\hline
(1) 1 TeV Higgs boson & 10.1 & 17   &  22.9 & 19 \\
(2) Chirally-coupled  Scalar & & & & \\
$m_S=1$ TeV, $\Gamma_S=0.35$ TeV &  7.1 & 12 & 11.3 & 9.3 \\
(3) O(2N) & 4.8 & 7.9 & 7.5 & 6.2 \\
(4) Chirally-coupled vector  & & & & \\
 $m_V=1.2$ TeV, $\Gamma_V=0.5$ TeV & 2.1 & 3.5 & 20.7 & 17 \\
(5) $m_V=1.5$ TeV, $\Gamma_V=0.6$ TeV & 0.36 & 0.6 & 4.3 & 3.6 \\
(6) LET & 2.6 & 4.3 & 1.7 & 1.4 \\
\hline
SM background &  3.7 & -  &  14.7 & - \\
\hline
\end{tabular}
\end{table}
%%%%%%%%%%%%%%%%%%%%%%%%%%%%%%%%%%%%%%%%%%%%%%%%%%%%%%%%%%%%%%%%%%%%%5
\newpage
\section*{Figures}

\begin{enumerate}

\item
\label{scheme}
Schematic diagram for longitudinal $W$-boson
scattering in $\gamma\gamma$ collisions.

\item
\label{feynman}
Typical Feynman diagrams contributing to the process $\gamma\gamma\to WWZZ$
and $WWWW$: (a) Higgs-boson exchange, (b) non-Higgs-boson  exchange.

\item
\label{cross}
Total cross sections of the process $\gamma\gamma\to W^+W^- ZZ$ versus the
center-of-mass energies $\sqrt{s_{\gamma\gamma}}$ of the $\gamma\gamma$ system
for $m_H=1.0$ (solid line) and 0.1 TeV (dashed line).

\item
\label{rapid}
Absolute rapidity distribution for the more-forward spectator $W$ boson in
the process $\gamma\gamma\to WWZZ$ with $m_H=1$ TeV and 0.1 TeV at
$\sqrt{s_{\gamma\gamma}}=2$ TeV.  Acceptance cuts are in Eqns.~(\ref{zz})
and (\ref{W}).

\item
\label{pt}
Transverse momentum $p_T$ distribution for the $Z$ boson with smaller $p_T$
in the process $\gamma\gamma \to WW ZZ$ at
$\sqrt{s_{\gamma\gamma}}=2$ TeV.  Acceptance cuts are in Eqn.~(\ref{zz}),
(\ref{W}), and (\ref{rap}).

\item
\label{summary1}
Summary curves for (a) $Z_L Z_L$ channel and (b) $W^+_L W^-_L$ channel:
cross sections of the  signal for various strong EWSB
models and the SM background.  The acceptance cuts on the strongly-scattered
$ZZ/WW$ pair are $m(ZZ/WW)>500$ GeV, $|y(Z/W)|<1.5$, and $p_T(Z/W)>250$ GeV;
while the acceptance cuts on the spectator $W$ bosons are $p_T(W_{\rm sp})>25$
 GeV and $|y(W_{\rm sp})|<3$, we also require  at least one forward spectator
$W$ boson in the rapidity region defined by $1.5 <|y(W_{\rm sp})| < 3$.
The 79\% tagging efficiency has been multiplied to the signal curves
to represent the effect of tagging the spectator $W$ bosons.
The models are (1) 1 TeV Higgs boson, (2) chirally-coupled scalar $m_S=1$ TeV
and $\Gamma_S=0.35$ TeV,  (3) O(2N) with $\Lambda=2$ TeV,
(4) chirally-coupled vector $m_V=1.2$ TeV and
$\Gamma_V=0.5$ TeV, (5) $m_V=1.5$ TeV and $\Gamma_V=0.6$ TeV, and (6) LET.
The SM background with $m_H=0.1$ TeV is indicated by (7).

\item
\label{summary2}
Summary curves: same as Fig.~\ref{summary1} but folded with the photon
spectrum using unpolarized laser and electron beams by laser backscattering.

\end{enumerate}

\end{document}